\documentclass[aps,pre,reprint,superscriptaddress]{revtex4-1}
\usepackage{graphicx}
\usepackage{dcolumn}
\usepackage{subfigure}

\begin{document}

\title{Hole-doping-induced half-metallic ferromagnetism in highly-air-stable PdSe$_2$ monolayer under uniaxial stress}

\author{Shi-Hao Zhang}
\affiliation{Beijing National Laboratory for Condensed Matter Physics, Institute of Physics, Chinese Academy of Sciences, Beijing 100190, China}
\affiliation{School of Physical Sciences, University of Chinese Academy of Sciences, Beijing 100190, China}
\author{Bang-Gui Liu}
\email[]{bgliu@iphy.ac.cn}
\affiliation{Beijing National Laboratory for Condensed Matter Physics, Institute of Physics, Chinese Academy of Sciences, Beijing 100190, China}
\affiliation{School of Physical Sciences, University of Chinese Academy of Sciences, Beijing 100190, China}

\date{\today}

\begin{abstract}
Two-dimensional (2D) high-temperature ferromagnetic materials are important for spintronic application. Fortunately, a highly-air-stable PdSe$_2$ monolayer semiconductor has been made through exfoliation from the layered bulk material. It is very highly desirable to realize robust ferromagnetism, even half-metallic ferromagnetism (100\% spin polarization), in such excellent nonmagnetic monolayer semiconductors. Here, the first-principles investigation shows that the PdSe$_2$ monolayer can be made to attain Stoner ferromagnetism with the maximal Curie temperature reaching to 800K, and the hole concentration threshold for ferromagnetism decreases with applied uniaxial stress. Furthermore, half-metallicity can be achieved in some hole concentration regions. For the strain of 10\% (uniaxial tensile stress of 4.4 N/m), the monolayer can attain half-metallic ferromagnetism up to 150 K. The magnetic anisotropic energy is suitable to not only stabilizing the 2D ferromagnetism but also realizing fast magnetization reversal. The magnetization can be also controlled by applying a transverse uniaxial stress. The highly-air-stable PdSe$_2$ monolayer, with these advantages, should be promising for spintronic applications.
\end{abstract}


\maketitle


\section{Introduction}

Recent experimental discovery of exfoliated two-dimensional ferromagnetic materials CrI$_3$~\cite{huang2017layer} and Cr$_2$Ge$_2$Te$_6$~\cite{gong2017discovery} inspires people to seek more two-dimensional realizable magnetic materials for device applications. Besides exfoliating two-dimensional magnetic van der Waals crystals, there are many methods to obtain the two-dimensional magnetism, such as electric field modulation~\cite{son2006half,BN}, strain engineering~\cite{zhou2012tensile,C6NR01333C}, nanoribbon edge modification~\cite{kan2008half}, surface adsorption~\cite{Hydrogenation,kan2010prediction,Uniformly}, transition-metal atom doping~\cite{C0CP02001J,Transition,li2017two}, defect engineering~\cite{zhang2013intrinsic,CPHC:CPHC201300097}, and so on. Carrier doping is always effective when there is high density of states near the Fermi level and the doped carrier can create itinerant ferromagnetism obeying Stoner's criterion $N(E_F)I > 1$, where $N(E_F)$ is the density of states at the Fermi energy in the nonmagnetic state and $I$ is Stoner parameter defined as $I = \Delta/M$ ($\Delta$ is the spin splitting energy and $M$ is the spin moment). The Stoner ferromagnetism has been predicted in the nonmagnetic two-dimensional materials GaSe~\cite{cao2015tunable}, InP$_3$~\cite{miao2017tunable}, PtSe$_2$~\cite{zulfiqar2016tunable}, C$_2$N~\cite{C7TC01399J} monolayer, phosphorene and arsenene~\cite{fu2017effects}.

Recently, PdSe$_2$ monolayer as a two-dimensional material  has been synthesised by exfoliating from bulk PdSe$_2$ crystals~\cite{PdSe2}. Very importantly, the semiconducting PdSe$_2$ monolayer has high air stability under ambient conditions, in contrast to the fast degradation of black phosphorus in air~\cite{li2014black,favron2015photooxidation}, which makes the PdSe$_2$ monolayer promising for electronic devices. Furthermore ultrathin PdSe$_2$ field-effect transistor with high mobility has been reported~\cite{ADMA}. Previous theoretical calculations showed that the monolayer has large Seebeck coefficients for both p- and n-type carriers when doping level is lower than 2$\times$10$^{13}$cm$^{-2}$~\cite{sun2015electronic}, but there is no magnetic exploration in the PdSe$_2$ monolayer, although it is highly desirable for spintronic applications based on highly-air-stable 2D materials.

The sharp peak in the density of states near the valence band maximum, always coexisting with large thermoelectric Seebeck coefficients~\cite{heremans2008enhancement}, implies that doping some carriers can move the Fermi level to the peaked density of states, and induce Stoner instability and then itinerant ferromagnetism in the PdSe$_2$ monolayer. In this work, our first-principles calculations show that doping hole into the PdSe$_2$ monolayer can create Stoner ferromagnetism and the Curie transition temperature can be far beyond room temperature. The magnetization, spin polarization energy,  magnetic anisotropic energy, and Curie temperature are systematically studied with different hole doping concentration under different uniaxial tensile stress. Stable half-metallic ferromagnetism, implying 100\% spin polarization at the Fermi level\cite{hm1983,hmlbg,Liu2005hm}, can be achieved by tuning the doped hole concentration and the applied uniaxial stress. More detailed results will be  presented in the following.

\begin{figure}[!htbp]
\includegraphics[width=0.35\textwidth]{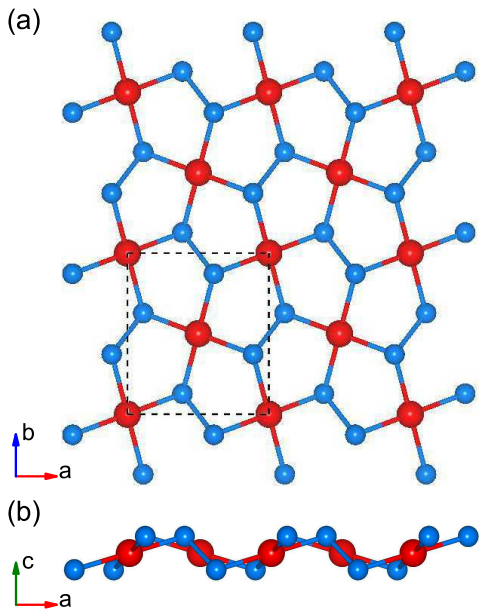}
\caption{~\label{fig1} The top view (a) and side view (b) of PdSe$_2$ monolayer. The red and blue balls represent Pd and Se atoms, respectively. The a, b, and c directions correspond to the y, x, and z axes, respectively.}
\end{figure}

\section{Computational methods}

Our first-principles calculations are performed with the Vienna Ab initio Simulation Package (VASP)~\cite{PhysRevB.47.558} with the projector-augmented wave (PAW) method~\cite{PhysRevB.50.17953}. The generalized gradient approximation (GGA) by Perdew, Burke, and Ernzerhof (PBE)~\cite{PhysRevLett.77.3865} is used as the exchange-correlation potential because the band gap calculated with PBE  is in good agreement with the experimental vaule~\cite{PdSe2}. The thickness of vacuum region is set as 20\AA{} to avoid any artificial interaction in the computational model. The cutoff energy is set to 500 eV, and the convergence standard is that the total energy difference between two successive steps is smaller than $10^{-6}$ eV. The structures are fully optimized to ensure all the Hellmann-Feynman forces on each atom are less than 0.01 eV/\AA{}. The hole doping is achieved by changing the total number of electrons of the unit cell, with a compensating jellium background of opposite charge added. Because the hole-doped magnetism is very sensitive to the sampling of densities of states (DOS), we carry out the Brillouin zone integration with a dense $\Gamma$-centered (41$\times$41$\times$1) Monkhorst-Pack grid~\cite{PhysRevB.13.5188}. When calculating the small magnetic anisotropic energy (MAE), the energy convergence criterion is promoted to $10^{-8}$ eV for achieving high accuracy.

\section{results and discussion}

\subsection{Intrinsic electronic structure}

\begin{figure*}[!htbp]
\includegraphics[width=0.96\textwidth]{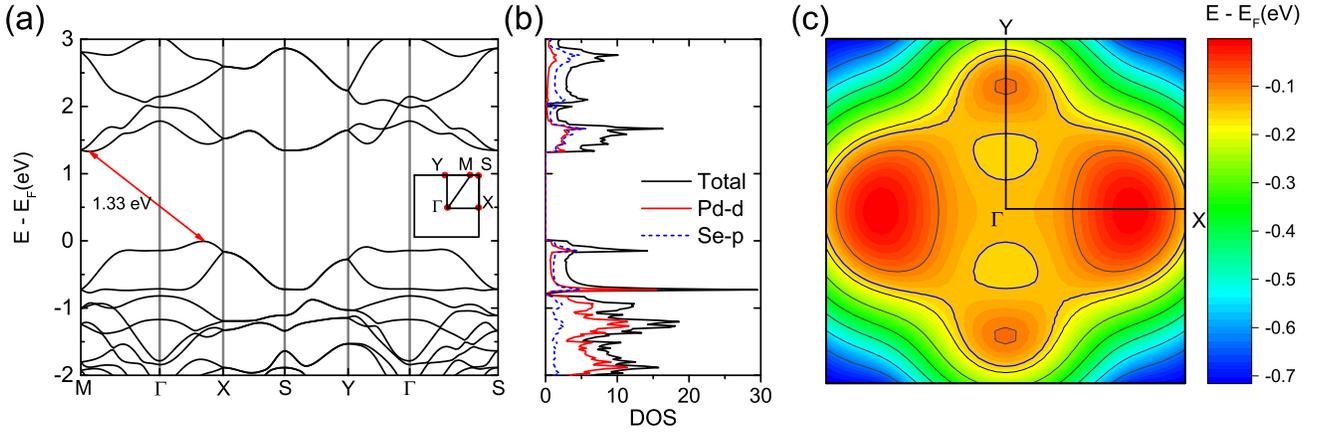}
\caption{~\label{fig2} The energy bands (a) and density of states (DOS) (b) of PdSe$_2$ monolayer, where the M point represents the (0.39, 0.5) point. (c) The energy distribution of the highest valence bands over the first Brillouin zone.}
\end{figure*}

The crystal structure of PdSe$_2$ monolayer~\cite{PdSe2} is shown in Fig.~\ref{fig1}. The unit cell of PdSe$_2$ monolayer (P2$_1$/c space group) has two palladium and four selenium atoms, with the lattice parameters $a = 5.74$ \AA{} (y-axis) and $b = 5.92$ \AA{} (x-axis). The monolayer shows puckering pentagonal ring which is like Cairo pentagonal tiling from the top view~\cite{PdSe2}. The energy bands and the density of states are showed in Fig.~\ref{fig2}. The monolayer shows semiconducting feature and has an indirect band gap of 1.33 eV which is agreement with the experiment~\cite{PdSe2}. The valence band maximum is located at (0.34, 0) and the conduction band minimum is situated on the (0.34, 0.44) point. In Fig.~\ref{fig2}(c), the energy values of the highest valence band over the entire Brillouin zone are presented. There are four maximum in the energy distribution, and  there is a flat band (within the blue contour)  and a sharp DOS peak at -0.15 eV (below the Fermi level). This DOS peak indicates a great probability for Stoner ferromagnetism induced by hole doping.

\begin{figure*}[!htbp]
\includegraphics[width=0.85\textwidth]{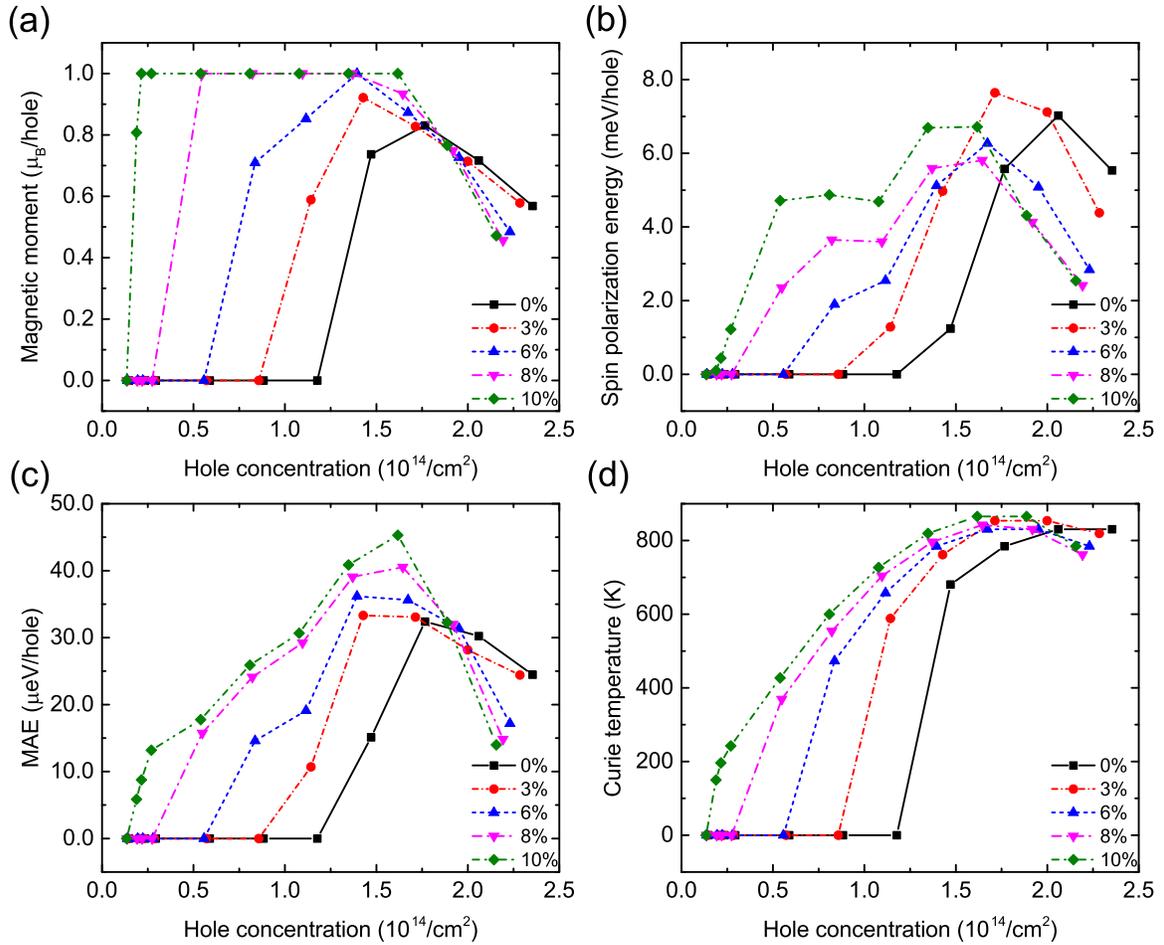}
\caption{~\label{fig3} The magnetic moment (a), spin polarization energy (b),  magnetic anisotropic energy (MAE) (c), and Curie temperature (d) versus hole doping concentration under different x-axis strain.}
\end{figure*}

\subsection{Hole doping and ferromagnetism}

We perform first-principles calculations about spin magnetic moment and spin polarization energy $E_{pol}$, the energy difference between the nonmagnetic and ferromagnetic states $E_{pol} = E_{non} - E_{fer}$). The calculated results are shown with solid lines in Fig.~\ref{fig3} (a,b). With introducing hole hoping, the Fermi level can touch the high density of states and then makes the strong on-site interactions between the opposite spins which induces the spin splitting. The calculated results shows that the system favors ferromagnetism under appropriate hole-doping levels because the energies of ferromagnetic states are lower than those of nonmagnetic cases. For PdSe$_2$ monolayer, ferromagnetism begins to occur when hole concentration is larger than 1.5$\times$10$^{14}$cm$^{-2}$ (0.25 hole per formula unit) and magnetic moment per hole becomes peaked (0.83 $\mu _B$ per hole) at the 1.8$\times$10$^{14}$cm$^{-2}$ (0.3 hole per formula unit) hole level. The positive spin polarization energy means stable magnetization and it reaches the maximum (7.0 meV per hole) at the 2.1$\times$10$^{14}$cm$^{-2}$ hole level, which is comparable to those of GaSe (3 meV per carrier)~\cite{cao2015tunable} and C$_2$N monolayer (8.5 meV per carrier)~\cite{C7TC01399J}.

Magnetic anisotropic energy (MAE) plays an important role in the two-dimensional stable long-range ferromagnetism~\cite{huang2017layer}. Defined as the total energy difference between the ferromagnetic configures along the out-of-plane (z) and the lowest in-plane (x or b) directions, MAE mainly originates from electronic contribution. The shape anisotropy caused by the dipole-dipole interactions can be neglected~\cite{PhysRevB.41.11919,PhysRevB.93.134407}. The calculated results are described with solid line in Fig.~\ref{fig3}(c). The maximum of MAE reaches at 32 $\mu$eV per hole for the hole concentration $1.8\times$10$^{14}$cm$^{-2}$, which is comparable to that of hole-doped phosporene~\cite{fu2017effects} and much larger than those of conventional transition metals: Fe, Co, and Ni (several $\mu$eV per atom)~\cite{PhysRevB.57.9557}. Compared to other two-dimensional materials with intrinsic magnetism, the hole-doped PdSe$_2$ monolayer has smaller MAE than CrXTe$_3$ (X=Si,Ge,Sn, 0.069-0.419 meV)~\cite{PhysRevB.92.035407}, CoBr$_2$ (2.6meV)~\cite{C7CP02158E}, and Fe$_2$Si (0.55--0.57meV)~\cite{acs.nanolett.6b04884} monolayers. This comparison indicates that the hole-doped PdSe$_2$ monolayers can be used as air-stable and appealing two-dimensional magnetic dynamic layers for fast spin dynamics.

\begin{figure*}[!htbp]
\includegraphics[width=0.85\textwidth]{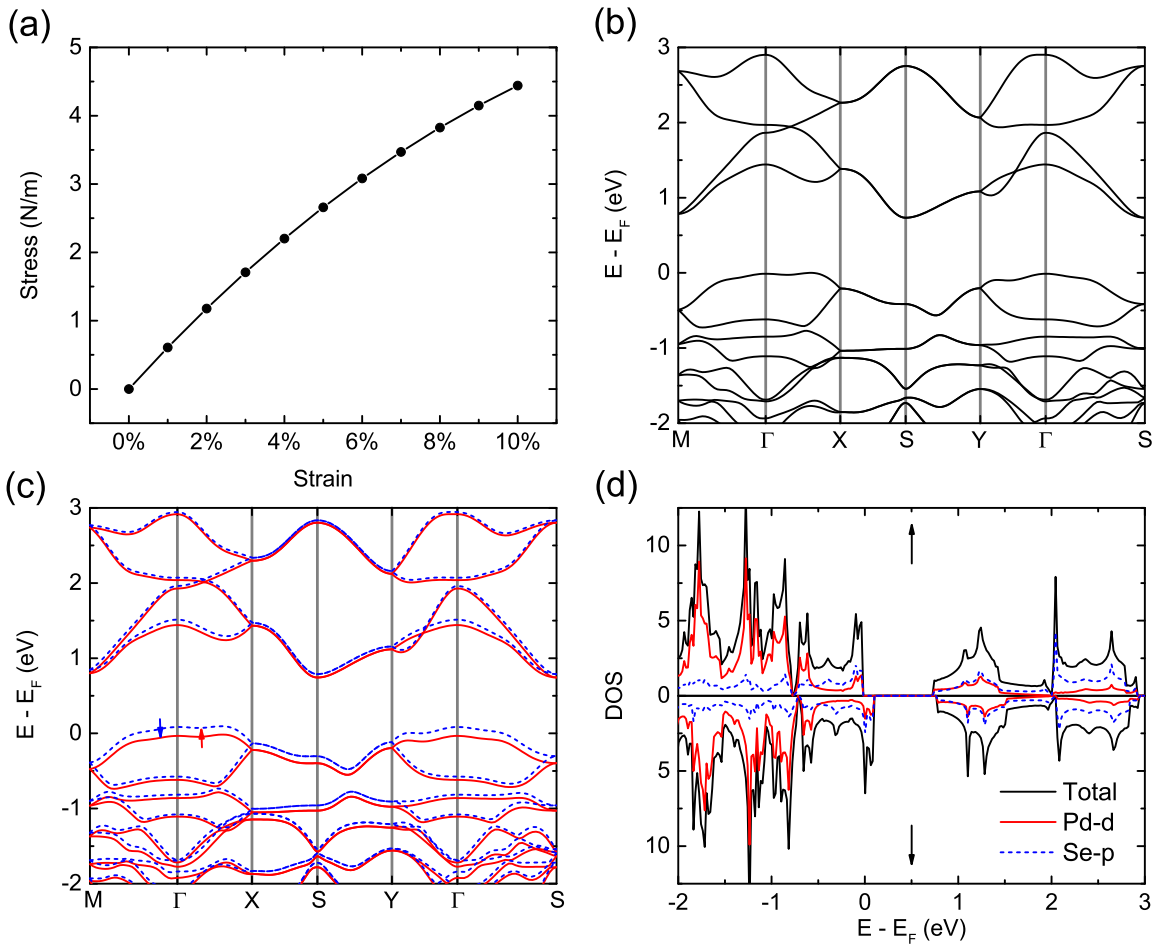}
\caption{~\label{fig4} (a) The uniaxial stress on the monolayer as a function of x-axis strain. (b) The energy bands of PdSe$_2$ monolayer under the x-axis strain of 10$\%$. The spin-polarized energy bands (c) and density of states (DOS) (d) of the monolayer with 0.2 holes per formula unit ($1.08\times 10^{14}$cm$^{-2}$) under the same strain.}
\end{figure*}

The Stoner magnetism originates from the strong exchange field in the system. When hole doping level is 1.8$\times$10$^{14}$cm$^{-2}$, the spin splitting energy at the $\Gamma$ point reaches 144 meV which corresponds to an effective Zeeman splitting from an external magnetic field of 1243 T. The exchange correlation induces the spontaneous magnetization obeying
\begin{eqnarray}
m&=&\frac{1}{N}\sum _{k}\left( \left \langle n_{k\sigma} \right \rangle - \left \langle n_{k\tilde{\sigma}} \right \rangle \right)\\
&=&\frac{1}{N}\sum _{k}\left\{ \frac{1}{e^{\beta (E_k-\Delta - \mu)}+1}-\frac{1}{e^{\beta (E_k+\Delta - \mu)}+1}\right\}
\end{eqnarray}
where $\beta$ is ($k_{B}T)^{-1}$, $\mu$ is the chemical potential, and $2\Delta$ is the spin splitting energy between two spin channels, $\sigma$ and $\tilde{\sigma}$.

Here we discuss how to estimate the Curie transition temperature of the itinerant ferromagnetism. We slowly increase the Gauss smearing factor to simulate the effect of increasing temperature $k_{B}T$. This will reduce the difference between $\left \langle n_{k\sigma} \right \rangle $ and $ \left \langle n_{k\tilde{\sigma}} \right \rangle$ in Eq. (1) and enlarge thermal excitations which weaken the magnetism and finally cause the Curie transition. By minimizing the free energy of the system at given temperatures, we can obtain the Curie temperature at the mean field level~\cite{cao2015tunable}. Calculated Curie transition temperature as a function of hole concentration is presented with solid line in Fig.~\ref{fig3}(d). The Curie temperatures are higher than 600 K when the hole concentration is larger than $1.4\times 10^{14}cm^{-2}$, which are much higher than those of GaSe monolayer (about 90 K)~\cite{cao2015tunable}. These results reveal that the stable Stoner ferromagnetism can be observed beyond the room temperature. The magnetization $M$ versus temperature $T$ for $1.8\times$10$^{14}$cm$^{-2}$ hole level shows a critical exponent of 1/2,  $\Delta M(T) \sim (T_c-T)^{1/2}$, where $T_{c}$ is Curie temperature. This is in agreement with the mean field model.

\subsection{Uniaxial stress and half-metallic ferromagnetism}

Tensile stress can always make the energy bands become more narrow (larger DOS) and thus make the Stoner ferromagnetism easier to occur. Actually, it was found that tensile strain decreases the critical value of hole concentration in the arsenene~\cite{fu2017effects} and PtSe$_2$ monolayer~\cite{zulfiqar2016tunable}. Due to the anisotropy between the x and y directions, uniaxial tensile stress is more accessible than biaxial stress. We apply a tensile stress along x-axis. As a result, it will cause a tensile strain in the x-axis and a compressive strain in the y-axis. The stress as a function of the x-axis strain is presented in Fig. 4(a). The strain can be applied by stretching a flexible substrate on which the 2D material is attaching, or electrically controlling by using a piezoelectric substrate~\cite{10.1021/nn401429w,akinwande2014two,10.1021/nn4024834}. It only needs 4.4 N/m to realize the x-axis strain of 10\% in the case of the PdSe$_2$ monolayer.

The calculated magnetic moment, spin polarization energy, magnetic anisotropic energy (MAE), and Curie temperature as functions of hole concentration for different x-axis strains are presented in Fig.~\ref{fig3}.
It is noted that the hole concentration threshold for ferromagnetism substantially decreases with increasing the x-axis strain, with the threshold being 1.5$\times$10$^{14}$cm$^{-2}$ for zero strain against 1.9$\times$10$^{13}$cm$^{-2}$ for 10$\%$. It should be noted that the hole concentration 2.0$\times$10$^{13}$cm$^{-2}$ corresponds to 0.04 holes per formula unit.

\begin{figure}[!htbp]
\includegraphics[width=0.45\textwidth]{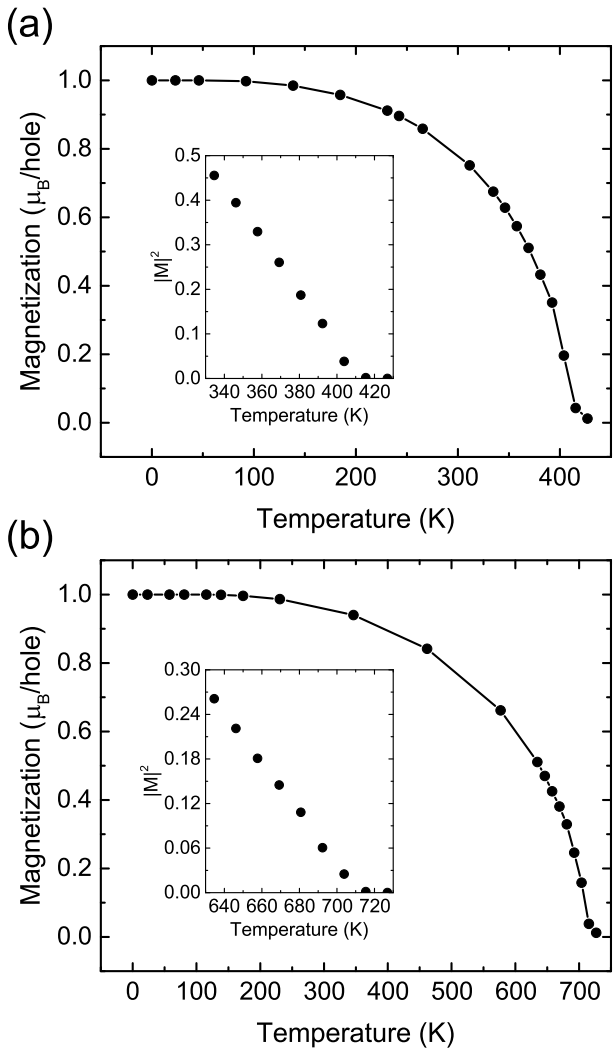}
\caption{~\label{fig5} The magnetizations versus temperature of PdSe$_2$ monolayer under x-axis strain 10\%, with the two hole concentrations of 5.4$\times$10$^{13}$cm$^{-2}$ (a) and 1.08$\times$10$^{14}$cm$^{-2}$ (b). Insets: The magnetization squared as a function of temperature near the Curie transition temperature for both of the cases.}
\end{figure}

Especially, half-metallic ferromagnetism begins to occur when the x-axis strain reaches to 6\%. It accompanies the saturated magnetic moment: $1\mu_B$ per hole. Keeping on increasing the strain, the doped hole concentration for half-metallicity develops into a plateau region. For example, it becomes $0.2\sim 1.6\times10^{14}$cm$^{-2}$ when the strain reaches 10$\%$. Because doped carrier concentrations were achieved to the order of 10$^{13}$cm$^{-2}$ in transition metal dichalcogenide monolayers by back-gate gating~\cite{mak2013tightly,zhang2014electrically} and to 10$^{14}$cm$^{-2}$ in graphene by ion liquid gating~\cite{PhysRevLett.105.256805,ye2011accessing}, the doped hole concentrations for half-metallic  ferromagnetism in the PdSe$_2$ monolayer should be accessible experimentally.

For comparison, we also present the energy bands of PdSe$_2$ monolayer under the x-axis strain of 10$\%$ without hole doping in Fig.~\ref{fig4}(b). The band width of the highest valence bands is 0.57 eV, smaller than that of the equilibrium state (0.72 eV). The energy bands and DOS of the monolayer under 10$\%$ with 0.2 holes per formula unit are showed in Fig.~\ref{fig4}(c,d). The DOS at the Fermi level, $N(E_F)$, for the nonmagnetic state with 0.2 holes per formula unit reaches 7.8/eV and the spin splitting energy near the Fermi level is 0.12 eV. Thus the Stoner parameter $I$ is 0.3 eV and the product $N(E_F)I$ equals 2.34, which satisfies the Stoner's criterion. It is clear that the energy bands and DOS show half-metallic ferromagnetism.

We also study the temperature dependence of the magnetization. For the strain of 10\%, we present the magnetization as a function of temperature for two hole concentrations of 5.4$\times$10$^{13}$cm$^{-2}$ and 1.08$\times$10$^{14}$cm$^{-2}$ in Fig. 5. These two hole concentrations are equivalent to 0.1 and 0.2 holes per formula unit, and the two Curie temperatures are higher than 400 K and 700 K, respectively. More importantly, the half-metallic ferromagnetism can persist up to 70K and 150 K, respectively. These imply that nearly 100\% spin polarization can be achieved and therefore the half-metallic PdSe$_2$ monolayer can be used for high-performance spintronic devices.

\subsection{Further discussions}

As for the spin polarization energy, the maximum is still approximately 7 meV per hole when the x-axis strain increases to 10$\%$, which proves the spin-polarized configures are still stable. The maximal Curie temperatures still remain approximately 800 K under different strains. In contrast, the maximum of MAE increases from 32 $\mu$eV per hole (0$\%$) to 45 $\mu$eV per hole (10$\%$), which implies that the uniaxial tensile stress can enhance the stability of the magnetization direction.

When a tensile stress is applied along the y-axis (the a direction), there will be a tensile strain in the y-axis and a compressive strain along the x-axis (the b direction). With 0.25 holes per formula unit doped into the monolayer, the magnetization along the x-axis can persist when a tensile y-axis strain up to 6$\%$ is applied. If the y-axis strain is larger than 6\%, the magnetization will switch from the x-axis to the y-axis. This is like ferroelectric polarization switching in GaTeCl monolayer\cite{GaTeCl}. Therefore, the stress can change the MAE of the monolayer, and can be used to control the magnetization direction of the hole-doped PdSe$_2$ monolayer.

\section{Conclusion}

The first-principles investigation has shown that when hole carriers rae doped into PdSe$_2$ monolayer, Stoner ferromagnetism can be induced and the maximal Curie temperature can reach to 800K. The hole concentration threshold for ferromagnetism decreases with applied stress (x-axis strain), reducing to 1.9$\times$10$^{13}$cm$^{-2}$ at the strain of 10\%. More importantly, half-metallicity can be formed in some hole concentration regions, in addition to the ferromagnetism. For the strain of 10\%, especially, when the hole doping concentration is in the range of $0.2\sim 1.6\times$10$^{14}$cm$^{-2}$, the monolayer can attain half-metallic ferromagnetism up to 150 K. These imply that 100\% spin polarization can be achieved in these hole concentration regions. A uniaxial tensile stress 4.4 N/m  can produce this large x-axis strain of 10$\%$. The magnetic anisotropic energy is suitable to stabilizing the two-dimensional ferromagnetism and ensuring fast magnetization reversal. The magnetization direction can be also controlled by applying a transverse uniaxial stress. The highly-air-stable PdSe$_2$ monolayer, with the high Curie temperature and robust half-metallic ferromagnetism, should be promising for spintronic applications.

\begin{acknowledgments}
This work is supported by the Nature Science Foundation of China (No.11574366), by the Strategic Priority Research Program of the Chinese Academy of Sciences (Grant No.XDB07000000), and by the Department of Science and Technology of China (Grant No.2016YFA0300701). The calculations were performed in the Milky Way \#2 supercomputer system at the National Supercomputer Center of Guangzhou, Guangzhou, China.
\end{acknowledgments}

%

\end{document}